\documentclass[aps,twocolumn,showpacs,preprintnumbers,amsmath,amssymb,superscriptaddress]
{revtex4}
\usepackage{graphicx}
\usepackage{dcolumn}% Align table columns on decimal point
\usepackage{bm}% bold math%
\usepackage{hyperref}

\begin{document}

\title
{Transverse force generated by an electric field and transverse
charge imbalance in spin-orbit coupled systems}

\author{Tsung-Wei Chen}
\email{twchen@phys.ntu.edu.tw} \affiliation{Department of Physics
and Center for Theoretical Sciences, National Taiwan University,
Taipei 106, Taiwan}

\author{Hsiu-Chuan Hsu}
\affiliation{Department of Physics and Center for Theoretical
Sciences, National Taiwan University, Taipei 106, Taiwan}

\author{Guang-Yu Guo}
\email{gyguo@phys.ntu.edu.tw} \affiliation{Department of Physics
and Center for Theoretical Sciences, National Taiwan University,
Taipei 106, Taiwan} \affiliation{Graduate Institute of Applied
Physics, National Chengchi University, Taipei 116, Taiwan}

\date{\today}

\begin{abstract}
We use linear response theory to study the transverse force generated by an
external electric field and hence possible charge Hall effect in
spin-orbit coupled systems. In addition to the Lorentz force that
is parallel to the electric field, we find that the transverse
force perpendicular to the applied electric field may not vanish
in a system with an anisotropic energy dispersion. Surprisingly,
in contrast to the previous results,
% of S. -Q. Shen, Phys. Rev. Lett. \textbf{95}, 187203 (2005),
the transverse force generated by the electric field does not
depend on the spin current, but in general, it is related to the
second derivative of energy dispersion only. The transverse force
always vanishes in the system with an isotropic energy dispersion.
However, the transverse force may also vanish in some systems with
an anisotropic energy dispersion such as the 2D k-cubic
Dresselhaus system. Furthermore, we find that the transverse force
does not vanish in the Rashba-Dresselhaus system. Therefore, the
non-vanishing transverse force acts as a driving force and results
in charge imbalance at the edges of the sample. This implies that
a non-zero Hall voltage can be detected in the absence of an
external magnetic field in anisotropic systems such as the
Rashba-Dresselhaus system. The estimated ratio of the Hall voltage
to the longitudinal voltage is $\sim 10^{-3}$. The disorder effect is also
considered in the study of the Rashba-Dresselhaus system. We find
that the transverse force vanishes in the presence of impurities
in this system because the vertex correction and the anomalous
velocity of the electron accidently cancel each other.
Nonetheless, we believe that the transverse charge imbalance
can be detected in the ballistic region by measuring the Hall voltage.
Our interesting prediction would stimulate measurements
of the Hall voltage in such spin-orbit coupled systems with an anisotropic
dispersion as the Rashba-Dresselhaus system in the near future.

\end{abstract}
 \maketitle

\section{Introduction}

The equation of motion of a charged particle in the presence of an
electromagnetic field is determined by the classical Lorentz
force~\cite{Jackson}. When only an electric field is present, the
charged particle is accelerated along the electric field
(longitudinal motion). In the presence of only a magnetic field,
the charged particle is deflected by the magnetic field when the
particle is in motion (cyclotron motion). In the 2D plane, the
combination of longitudinal and cyclotron motion would result in
the appearance of the classical charge-Hall effect~\cite{Ashcroft}
when the applied electric field and magnetic field are parallel
and perpendicular to the plane, respectively. In the absence of a
magnetic field, the charge Hall effect would disappear in the
classical regime.

When the spin degree of freedom is considered, the
charged-particle motion further changes; in particular, the
orbital motion now depends on the spin direction (spin-orbit
interaction). The anomalous motion of particles that is induced by
spin-orbit interaction has attracted much attention. The
semiclassical force exerted on a spin-1/2 electron in the presence
of an electromagnetic field has been investigated~\cite{Ryu96}.
The force acts on the spin according to SU(2) non-abelian gauge
theory because the spin-orbit interaction
$(\boldsymbol{\sigma}\times\mathbf{E})\cdot\mathbf{p}$, which is
taken to be the kinetic momentum, plays the role of the gauge
field~\cite{Ryu96}. It has been shown that the equation of motion
of an charged particle in a crystal environment can be altered by
the Berry curvature of band structure~\cite{MCChang96, MCChang08}
when the spin-orbit interaction is considered:
$\dot{\mathbf{x}}=\frac{\partial\epsilon_n(\mathbf{k})}{\hbar\partial\mathbf{k}}-\dot{\mathbf{k}}\times\mathbf{B}_n(\mathbf{k})$.
The Berry curvature correction
($-\dot{\mathbf{k}}\times\mathbf{B}_n(\mathbf{k})$) can also be
derived by using the Feynmann path integral method~\cite{Koiz02}.
The Berry curvature plays the role of an effective magnetic field
and thus results in the anomalous deviation of the classical
trajectory. In this context, an interesting phenomenon was
observed in the study described in Ref.~\cite{Mur03}: in the
adiabatic approximation, the transverse velocity due to Berry
curvature exists even in the absence of an external magnetic
field. It was further shown that in the adiabatic sense, the
non-vanishing Berry curvature near the Brillouin zone center of a
hole-doped (p-type) semiconductor would lead to intrinsic
spin-Hall effect~\cite{Mur03}. This discovery led to an
interesting result: the magnetic monopole would exist in the
crystal momentum space of solids~\cite{Fang03}. Furthermore,
the concept of the anomalous velocity due to Berry curvature
has been used in ab initio relativistic band structure calculations
of the intrinsic spin Hall conductivity in p-type semiconductors
\cite{Guo05} and also Pt metal~\cite{Guo08}.
%The intrinsic
%spin-Hall conductivity for hole-doped GaAs semiconductor was also
%studied in Ref.~\cite{Mur04}, where the conserved spin currents in
%both heavy- and light-hole bands were established by using SO(4)
%symmetry. On the other hand, in \cite{Guo05}, {\it ab inito}
%relativistic band structure calculations show that the orbital
%angular momentum Hall conductivity in p-type semiconductors is one
%order of magnitude smaller than the spin Hall conductivity,
%indicating no cancellation between the spin and orbital angular
%momentum Hall effects in bulk semiconductors. Furthermore, the
%large spin-Hall effect in Pt metal at room
%temperature~\cite{Kimu07} was also theoretically investigated. It
%was attributed to be an intrinisc one due to the band
%anti-crossings near the Fermi level at the $L$ and $X$ symmetry
%points in the Brillouin zone~\cite{Guo08}.

Due to the existence of Berry curvature~\cite{Mur03}, the
intrinsic force of the k-linear Rashba system is studied in
semiconductor wires in ballistic regime~\cite{Nik05}. The force
was semiclassically defined as the time derivative of the kinetic
momentum in the Heisenberg picture. It was found that the
deflection of the center of spin-polarized wave packet in the
transverse direction can be explained by the existence of the
intrinsic force~\cite{Nik05}. The concept of intrinsic force due
to the spin-orbit interaction in semiconductor is further
generalized to the non-relativistic limit of the Dirac
equation~\cite{Shen05} in which the spin-orbit interaction is
still linear in momentum. In the non-relativistic limit of the
Dirac equation, it has been shown that a semiclassical force is
exerted on the moving particle with spin~\cite{Shen05}. For the 2D
system, the effective Hamiltonian describing the
Rashba-Dresselhaus system is found to have the structure of the
SU(2)$\times$ U(1) gauge field~\cite{Jin06}. The transverse force
derived from the corresponding four-force in the
Rashba-Dresselhaus system agrees with the result in
Ref.~\cite{Shen05}. Recently, the inverse spin-Hall effect driven
by the spin motive force in a ferromagnetic conductor has been
studied~\cite{Shi09}. The Hall angle in this case is shown to be
greater than that in the anomalous Hall effect since skew and side
jump scattering are considered~\cite{Shi09}.

However, the force in the linear response to an external electric
field in a spin-orbit coupled system has not yet been studied.
This is very important in the study of forces generated by
external fields because the force operator is significant only
when the expectation value is taken into account. In the present
paper, we study the force in spin-orbit coupled systems with the
generic two-band effective Hamiltonian [Eq. (\ref{Ham})] by using
linear response theory. The force is semiclassically defined as
the time derivative of the kinetic momentum $m\mathbf{v}$. The
velocity $\mathbf{v}$ can be separated into two terms,
$\mathbf{v}=\mathbf{v}_f+\mathbf{v}_a$, where
$\mathbf{v}_f=\mathbf{p}/m$ and $\mathbf{v}_a$ is the anomalous
velocity due to the spin-orbit interaction. The spin-orbit
interaction induced force (SOIF) originates from the time
derivative of $m\mathbf{v}_a$. In this study, we shall focus on
the SOIF in the linear response to an external electric field. Our
main result is that the SOIF depends only on the second derivative
of energy dispersion and not on the spin~\cite{Mur03} or spin
current~\cite{Shen05}. We provide an analytic expression for the
SOIF and show that the non-vanishing transverse SOIF may exist
when the system is not rotationally invariant. This also implies
that the charge imbalance at the edges of the sample may occur in
a spin-orbit coupled system because the non-vanishing transverse
force acts as a driving force in the system.

Our present paper is organized as follows. In Sec. II, we derive
the force from the generic effective Hamiltonian by using the
semiclassical definition, that is, the time derivative of kinetic
momentum $m\mathbf{v}$. In Sec. III, we use linear response theory
to investigate the properties of the force under a weak and
constant electric field. In Sec. IV, we study the
Rashba-Dresselhaus system and also consider the presence of weak
disorder. Our conclusions are presented in Sec. V.

%Appendix A and B of this paper outline our derivations of linear
%response of intrinsic force to the electric field and the vertex
%correction calculations.

%This provides an alternative way to detect whether the system we
%considered is isotropic at some carrier concentration. We then
%calculate the average SOIF for Rashba-Dresselhaus system, and find
%that the transverse SOIF is in general non zero. We also discuss
%an experiment to test our prediction.

\section{Force generated by spin-orbit interaction}

In this section, we will derive the intrinsic force in the linear
response to an electric field. We assume that the effective
Hamiltonian can be partitioned into two terms: a kinetic energy
term and spin-orbit interaction term. In the absence of spin-orbit
interaction, the Hamiltonian would simplify to the free electron
(hole) model. In this case, we consider the following effective
Hamiltonian that includes the spin-orbit interaction:
\begin{equation}\label{Ham}
H_0=\frac{\hbar^2\mathbf{k}^2}{2m}+h_{so},
\end{equation}
where the first term of Eq. (\ref{Ham}) represents the kinetic
energy for a free effective mass $m$ and $h_{so}$ represents the
spin-orbit interaction. We do not consider the external magnetic
field, and thus, the time reversal symmetry is preserved. By
spectral decomposition, the spin-orbit interaction $h_{so}$ can be
written as
\begin{equation}\label{HamSO}
\begin{split}
h_{so}&=\sum_{n}\Delta_n(\mathbf{k})\mathcal{P}_{nk}\\
\mathcal{P}_{nk}&=|n\mathbf{k}\rangle\langle n\mathbf{k}|,
\end{split}
\end{equation}
against the background that the eigenenergy of the Hamiltonian
$H_0$ is
$E_{n\mathbf{k}}=\frac{\hbar^2\mathbf{k}^2}{2m}+\Delta_n(\mathbf{k})$;
in other words,
$H_0|n\mathbf{k}\rangle=E_{n\mathbf{k}}|n\mathbf{k}\rangle$.
$\mathcal{P}_{nk}$ is the projection operator that satisfies the
relation
$\mathcal{P}_{nk}\mathcal{P}_{mk}=\mathcal{P}_{nk}\delta_{mn}$.
The index $n$ denotes the band index, and $\Delta_n(\mathbf{k})$
is the energy dispersion of spin-splitting for the $n$-th band. It
must be emphasized that the dispersion $\Delta_n(\mathbf{k})$ is
not limited to the odd function of $\mathbf{k}$ in which the
system lacks an inversion center. For GaAs, the parity
selection rule does not allow the appearance of a $k$ linear term
for the hole band, and the spin-orbit interaction term of the
Luttinger Hamiltonian is quadratic in $k$. For a spherical
approximation of the Luttinger Hamiltonian (see Ref.~\cite{Mur03})
$H_0=\frac{\hbar^2}{2m_0}[(\gamma_1+\frac{5}{2}\gamma_2)k^2-2\gamma_2(\mathbf{k}\cdot\mathbf{S})^2]$,
we have
$h_{so}=\frac{\hbar^2}{2m_0}\gamma_2[\frac{5}{2}k^2-2(\mathbf{k}\cdot\mathbf{S})^2]$
for the spin-orbit interaction, $m=m_0/\gamma_1$ for the free
effective mass, and
$\Delta_{\pm\frac{3}{2}}(\mathbf{k})=-2\gamma_2\frac{\hbar^2k^2}{2m_0}$
and
$\Delta_{\pm\frac{1}{2}}(\mathbf{k})=+2\gamma_2\frac{\hbar^2k^2}{2m_0}$
represent the spin-splittings of heavy and light holes,
respectively. On the other hand, for a 2D system, the spin
splitting can be simply written as
$\Delta_{n}(\mathbf{k})=-n\Delta(\mathbf{k})$, which indicates
that the magnitude of spin-splitting in each band is the same. In
the following discussions, we will show that only a $k$-linear
term can reproduce the result given in Ref.~\cite{Shen05}. For
terms containing higher orders of $k$, in general, the force
operator is not related to the spin current. We will show that in
the 2D system, the force is actually related by the Berry vector
potential. The common 2D effective Hamiltonian that includes
spin-orbit interaction can also be written in the form of Eq.
(\ref{Ham}) (see also Ref.~\cite{TWChen08}). It should be pointed
out that the knowledge of explicit form of the energy dispersion
$\Delta(\mathbf{k})$ and eigenfunction $|n\mathbf{k}\rangle$ is
not necessary in the following derivations. We will derive an
analytic expression for the linear response of the semiclassical
force.

In the presence of a constant electric field, the potential can be
written as $-q\mathbf{E}\cdot\mathbf{x}$, and we have
\begin{equation}
H=H_0-q\mathbf{E}\cdot\mathbf{x},
\end{equation}
where $q$ is the charge of the particle ($q=-|e|$ for an
electron). The notation $\mathbf{x}$ refers to the gauge invariant
position operator~\cite{Chen06}. The diagonal terms of the gauge
invariant position operator vanish, and the off-diagonal terms is
equivalent to the replacement of $\mathbf{x}$ with
$i\frac{\partial}{\partial\mathbf{k}}$. The force in the following
derivation is defined as the time derivative of kinetic momentum
$m\mathbf{v}$, and we have
\begin{equation}
\mathbf{F}=m\frac{d\mathbf{v}}{dt}=\frac{m}{i\hbar}[\mathbf{v},H],
\end{equation}
where the velocity operator is defined as
$\mathbf{v}=\frac{1}{i\hbar}[\mathbf{x},H]=\frac{\partial
H}{\partial\mathbf{p}}$, where the gauge invariant position
operator does not change the original commutator~\cite{Chen06}.
The velocity operator is composed of two terms: the free electron
(hole) velocity $\mathbf{v}_f=\mathbf{p}/m$ and the anomalous
velocity $\mathbf{v}_{a}=\partial h_{so}/\partial\mathbf{p}$,
which is induced by the spin-orbit interaction. By the use of
equation $\mathbf{v}=\mathbf{v}_f+\mathbf{v}_a$, the
semi-classical force
$\mathbf{F}=m\frac{d\mathbf{v}}{dt}=\frac{m}{i\hbar}[\mathbf{v},H]$
is composed of two terms:
\begin{equation}\label{TForce}
\mathbf{F}=\mathbf{F}^L+\mathbf{F}^{SOI},
\end{equation}
where $\mathbf{F}^{L}=\frac{m}{i\hbar}[\mathbf{v}_f,H]$ can be
written as (the Einstein summation convention is used)
\begin{equation}\label{Lorenz}
F^L_i=q\delta_{ij}E_j,
\end{equation}
which is the Lorentz force.
$\mathbf{F}^{SOI}=\frac{m}{i\hbar}[\mathbf{v}_a,H]$ is the SOIF
operator. For convenience in the following discussions, the SOIF
operator is rewritten as the sum of two terms:
$\mathbf{F}^{SOI}=\mathbf{F}^{E}+\mathbf{F}^O$. The force
$\mathbf{F}^{E}=\frac{m}{i\hbar}[\frac{\partial
h_{so}}{\hbar\partial\mathbf{k}},-q\mathbf{E}\cdot\mathbf{x}]$
depends on the anisotropic properties of energy dispersion, and it
can be shown that
\begin{equation}\label{SOIF}
F^{E}_i=\frac{qm}{\hbar^2}\frac{\partial^2h_{so}}{\partial
k_i\partial k_j}E_j.
\end{equation}
The force $\mathbf{F}^{O}$ originates from the commutator of the
anomalous component of the velocity and the spin orbit interaction
of the Hamiltonian, namely, $[\mathbf{v}_{a},h_{so}]$, and it can
be written as
\begin{equation}\label{SOIF2}
F^{O}_i=\frac{m}{i\hbar^2}[\frac{\partial h_{so}}{\partial
k_i},h_{so}].
\end{equation}
In general, Eq. (\ref{SOIF2}) has a complicated form, but in a 2D
spin-orbit coupled system, it can be shown that the force
$F^{O}_{i}$ [Eq. (\ref{SOIF2})] is related to the Berry vector
potential. This can be seen as follows. For 2D systems, the
Hamiltonian can generally be written as
$H_0=\frac{\hbar^2k^2}{2m}+A(\mathbf{k})\sigma_x-B(\mathbf{k})\sigma_y$,
with energy dispersion
$E_{nk}=\frac{\hbar^2k^2}{2m}-n\sqrt{A^2+B^2}$. It can be shown
that
\begin{equation}\label{SCF}
\frac{m}{i\hbar^2}[\frac{\partial h_{so}}{\partial
k_i},h_{so}]=-\frac{m\Omega^2}{2}\left(\frac{\partial\theta}{\partial
k_i}\right)\sigma_z,
\end{equation}
where $\theta(\mathbf{k})$ is defined as
$\theta=\tan^{-1}\left(\frac{A(\mathbf{k})}{B(\mathbf{k})}\right)$
and $\Omega=\frac{2\Delta}{\hbar}$. For an appropriate choice of
phase of eigenstate, one can obtain $\langle
n\mathbf{k}|(-i)\frac{\partial}{\partial\mathbf{k}}|n\mathbf{k}\rangle=\frac{1}{2}\frac{\partial\theta}{\partial\mathbf{k}}$.
In Ref.~\cite{BZhou06}, it was shown that the spin-Hall
conductivity can be expressed in terms of anti-commutator of
Eq.~(\ref{SOIF2}) and spin current operator.

On the other hand, it is interesting to note that for k-linear
systems such as the Rashba-Dresselhaus system, Eq. (\ref{SOIF2})
can be further written as
$\frac{4m^2}{\hbar^4}(\alpha^2-\beta^2)(\mathbf{J}^{s_z}\times\hat{e}_z)$,
where $\mathbf{J}^{s_z}=\frac{1}{2}\{\mathbf{v},s_z\}$ is the
conventional definition of spin current, as shown in
Ref.~\cite{Shen05}, where the spin force corresponding to
$\mathbf{F}^{O}$ is derived by considering the non-relativistic
limit of the Dirac equation. However, the force $\mathbf{F}^{O}$
is not always related to the conventional spin current. For
example, for the 2D k-cubic Dresselhaus
system~\cite{Malshu05,TWChen08}, where the spin-orbit interaction
is $h_{so}=\beta_D(k_xk_y^2\sigma_x-k_yk_x^2\sigma_y) $, and the
energy dispersion is $\Delta(\mathbf{k})=\beta_Dkk_xk_y$, we have
$\partial\theta/\partial k_x=-\frac{k_y}{k^2}$. Therefore, Eq.
(\ref{SOIF2}) can be written as
$\frac{4m\beta_D}{\hbar^2}k_x^2k_y^2(k_x\sigma_z)$, where the
spin-Hall current can not be determined.

\section{Linear response to weak electric field}
Using Ehrenfest's theorem, the time derivative of the expectation
value of the velocity is equal to the commutator of velocity and
system Hamiltonian, i.e.,
\begin{equation}\label{EhrenF}
\frac{\partial}{\partial
t}\left[\Psi^{\dag}(t)\mathbf{v}\Psi(t)\right]=Re\left[\Psi^{\dag}(t)\frac{1}{i\hbar}[\mathbf{\mathbf{v}},H]\Psi(t)\right],
\end{equation}
where $Re[\cdots]$ represents the real part of $[\cdots]$. The
state vector $\Psi(t)$ satisfies the Shr$\ddot{o}$dinger equation
$H\Psi=i\hbar\frac{\partial\Psi}{\partial t}$ and
$H=H_0-q\mathbf{E}\cdot\mathbf{x}$ ($H_0$ is given by Eq.
(\ref{Ham})). In the derivation of Eq. (\ref{EhrenF}), the fact
that $\{v_i,H\}$ is a hermitian operator is used. Accordingly, the
left-hand side is defined as the force in the system, and the
right-hand side of Eq. (\ref{EhrenF}) can be expanded up to the
first order of the applied electric field. In linear response
theory, the expectation value of the operator $\mathcal{O}$ can be
written as
\begin{equation}\label{LRT1}
\begin{split}
\langle\mathcal{O}\rangle&=\frac{1}{\mathcal{V}}\sum_{n\mathbf{k}}f_{n\mathbf{k}}\langle
n\mathbf{k}|\mathcal{O}|n\mathbf{k}\rangle\\
&~~+\frac{1}{\mathcal{V}}\sum_{n\mathbf{k}}f_{n\mathbf{k}}2Re[\langle
n\mathbf{k}|\mathcal{O}|n\mathbf{k}\rangle^{1}]+o(E^2),
\end{split}
\end{equation}
where $f_{n\mathbf{k}}$ is the Fermi-Dirac distribution and
$\mathcal{V}$ is the volume of the system. In the right-hand side
of the equality, the first term of Eq. (\ref{LRT1}) gives the
expectation value of the unperturbed wave function and the second
term of Eq. (\ref{LRT1}) represents the linear response to the
weak electric field. The perturbed wave function to first order of
perturbation $V$, denoted as $|n\mathbf{k}\rangle^1$, is given by
\begin{equation}\label{LRT2}
|n\mathbf{k}\rangle^{1}=\sum_{n'(\neq
n)}\frac{|n'\mathbf{k}\rangle\langle
n'\mathbf{k}|V|n\mathbf{k}\rangle}{E_{n\mathbf{k}}-E_{n'\mathbf{k}}}.
\end{equation}
In our case, the perturbation is $V=-q\mathbf{E}\cdot\mathbf{x}$.
It must be emphasized that the gauge invariant position operator
does not change the matrix element of $\langle
n'\mathbf{k}|V|n\mathbf{k}\rangle$ because only the interband
transition is considered in Eq. (\ref{LRT2})~\cite{Chen06}.

In the following calculation, the expectation value is evaluated
to the first order in the electric field. For the Lorentz force
$\mathbf{F}^{L}$ [Eq. (\ref{Lorenz})], we have (to the first order
in the electric field)
\begin{equation}
\langle F_i^{L}\rangle=qn_c\delta_{ij}E_j,
\end{equation}
where $n_c$ is the carrier concentration. The Lorentz force acts
only in longitudinal direction. It is important to note that the
Lorentz force is derived from the commutator of the kinetic
momentum $\frac{\mathbf{p}}{m}$ and potential
$V=-q\mathbf{E}\cdot\mathbf{x}$. In single-layer graphene and
bilayer graphene, there is no kinetic energy in effective
tight-binding Hamiltonians~\cite{Cserti06}, and thus, the charged
particle does not experience the Lorentz force in either of the
systems. In calculating the linear response of $\mathbf{F}^{E}$
[Eq. (\ref{SOIF})] and $\mathbf{F}^O$ [Eq. (\ref{SOIF2})], we note
that Eq. (\ref{SOIF}) includes the effect of electric field but
Eq. (\ref{SOIF2}) does not. By using Eq. (\ref{HamSO}) and the
first term of Eq. (\ref{LRT1}), it can be shown that
\begin{equation}\label{hso1}
\begin{split}
\langle F^E_i\rangle&=\frac{1}{\mathcal{V}}\sum_{n\mathbf{k}}
f_{n\mathbf{k}}\langle
n\mathbf{k}|\frac{qmE_j}{\hbar^2}\frac{\partial^2h_{so}}{\partial
k_i\partial
k_j}|n\mathbf{k}\rangle\\
&=\frac{1}{\mathcal{V}}\sum_{n\mathbf{k}}
f_{n\mathbf{k}}\left[\frac{\partial^2\Delta_n(\mathbf{k})}{\partial
k_i\partial
k_j}\frac{qm}{\hbar^2}E_j+\frac{qm}{\hbar^2}\Gamma^{n}_{ij}E_j\right],
\end{split}
\end{equation}
where
\begin{equation}\label{hso2}
\begin{split}
\Gamma^n_{ij}=&~-2\Delta_{n}
Re\left[\langle\frac{\partial(n\mathbf{k})}{\partial
k_i}|\frac{\partial(n\mathbf{k})}{\partial
k_j}\rangle\right]\\
&+2Re\left[\langle\frac{\partial(n\mathbf{k})}{\partial
k_i}|h_{so}|\frac{\partial(n\mathbf{k})}{\partial
k_j}\rangle\right].
\end{split}
\end{equation}
The calculation of second term of Eq. (\ref{LRT1}) for $F^E_i$ is
not necessary because it gives a second order of an electric
field. The linear response of Eq. (\ref{SOIF2}) is obtained from
the expectation value $2Re\langle
n\mathbf{k}|\frac{1}{i}[\frac{\partial h_{so}}{\partial
k_i},h_{so}]|n\mathbf{k}\rangle^1$ because the first term of Eq.
(\ref{LRT1}) vanishes, i.e., it can be shown that $\langle
n\mathbf{k}|\frac{1}{i}[\frac{\partial h_{so}}{\partial
k_i},h_{so}]|n\mathbf{k}\rangle=0$. For interband transition, the
gauge invariant position operator $\mathbf{x}$ can be replaced by
$i\frac{\partial}{\partial\mathbf{k}}$, and we have
\begin{equation}\label{hso3}
\begin{split}
&2Re\langle n\mathbf{k}|\frac{1}{i}[\frac{\partial
h_{so}}{\partial k_i},h_{so}]|n\mathbf{k}\rangle^1\\
&=2qE_j\sum_{\ell(\ell\neq n)}(\Delta_n-\Delta_{\ell})
Re\left[\langle\frac{\partial(n\mathbf{k})}{\partial
k_i}|\ell\mathbf{k}\rangle\langle\ell\mathbf{k}|\frac{\partial(n\mathbf{k})}{\partial
k_j}\rangle\right]\\
&=-\frac{qm}{\hbar^2}\Gamma^{n}_{ij}E_j.
\end{split}
\end{equation}
We note that the restriction $\ell\neq n$ in the first equality of
Eq. (\ref{hso3}) is not necessary because the term corresponding
to $\ell=n$ is zero. Furthermore, the second equality is carried
out by using the fact that
$h_{so}=\sum_{\ell}\Delta_{\ell}|\ell\mathbf{k}\rangle\langle\ell\mathbf{k}|$
and $\sum_{\ell}|\ell\mathbf{k}\rangle\langle\ell\mathbf{k}|=1$.
We obtain
\begin{equation}\label{hso4}
\begin{split}
\langle F^O_i\rangle&=\frac{1}{\mathcal{V}}\sum_{n\mathbf{k}}
f_{n\mathbf{k}}2Re\langle
n\mathbf{k}|\frac{m}{i\hbar^2}[\frac{\partial h_{so}}{\partial
k_i},h_{so}]|n\mathbf{k}\rangle^1\\
&=\frac{1}{\mathcal{V}}\sum_{n\mathbf{k}}
f_{n\mathbf{k}}\left(-\frac{qm}{\hbar^2}\Gamma^{n}_{ij}E_j\right).
\end{split}
\end{equation}
It is interesting to note that Eq. (\ref{hso4}) exactly cancels
the second term of Eq. (\ref{hso1}). This implies that in fact,
the force $\langle\mathbf{F}^{O}\rangle$ does not exist in the
linear response to a static electric field. The resulting mean
SOIF $\langle F^{SOI}_{i}\rangle=\langle F^{E}_i+F^{O}_i\rangle$
is (the Einstein summation convention is used)
\begin{equation}\label{hsoEO}
\langle F^{SOI}_{i}\rangle\equiv\langle
F^{E}_{i}+F^{O}_i\rangle=qK_{ij}E_j,
\end{equation}
where
\begin{equation}\label{Kij}
K_{ij}=\frac{m}{\hbar^2\mathcal{V}}\sum_{n\mathbf{k}}f_{n\mathbf{k}}\frac{\partial\Delta_{n}(\mathbf{k})}{\partial
k_i\partial k_j}.
\end{equation}
Therefore, in the system with spin-orbit interaction, the particle
would experience two kinds of forces when only an electric field
is present:
\begin{equation}\label{TotalF}
\langle F_{i}\rangle=q(n_c\delta_{ij}+K_{ij})E_j,
\end{equation}
where $K_{ij}$ is given in Eq. (\ref{Kij}). The first term
represents the Lorentz force and the second term represents the
force generated by the spin-orbit interaction. Physically,
$K_{ij}$ is the inverse effective mass matrix associated with the
spin-orbit interaction. The anisotropic effective mass can be
defined as $\frac{dv_i}{dt}=[\frac{1}{M}]_{ij}qn_cE_j$, where
$[\frac{1}{M}]_{ij}=\frac{1}{m}(\delta_{ij}+\frac{K_{ij}}{n_c})$.
If all the off-diagonal components of $K_{ij}$ vanish in some
system, then the diagonal components of the effective mass are
equal to the free effective mass. This implies that the transverse
effective mass is zero, and thus, the non-zero transverse SOIF is
forbidden in this system. However, for the Rashba-Dresselhaus
system, we find that the transverse SOIF does not vanish. We will
discuss the Rashba-Dresselhaus system in the next section.

We note that the mean SOIF $\langle\mathbf{F}^{SOI}\rangle$
depends on the second derivative of the energy dispersion.
Unlike the force
$\frac{m}{i\hbar^2}[\frac{\partial h_{so}}{\partial k_i},h_{so}]$
that is related to the conventional definition of spin current in
k-linear systems, the mean SOIF $\langle\mathbf{F}^{SOI}\rangle$
is generally independent of the definition of spin current. We
also note that the mean SOIF $\langle\mathbf{F}^{SOI}\rangle$ is
analogous to the semiclassical force acting on a band electron (or
hole), which can be written as
$\hbar\frac{d\mathbf{k}}{dt}=q\mathbf{E}$ and
$\mathbf{v}=\frac{\partial
E_{n\mathbf{k}}}{\partial\hbar\mathbf{k}}$ (see
Ref.~\cite{Ashcroft}). From these two equations, we have
$\frac{dv_i}{dt}=\frac{\partial^2 E_{n\mathbf{k}}}{\partial
k_i\partial k_j}\frac{dk_j}{\hbar dt}=\frac{\partial
E_{nk}}{\hbar^2\partial k_i\partial k_j}qE_j$; this is in
agreement with the result of the linear response calculation.
However, it must be pointed out that the force generated by
spin-orbit interaction has not yet been studied. An important
indication of our result is that the inclusion of the spin-orbit
interaction does not change the form of the semiclassical force.

We now turn to the discussion of the k-linear system where
$\mathbf{F}^{E}=0$. We note that in the k-linear system ($h_{so}$
contains only k-linear terms), we must have $\frac{\partial^2
h_{so}}{\partial k_i\partial k_j}=0$ and thus $\mathbf{F}^{E}=0$.
This means that $\langle\mathbf{F}^{E}\rangle$ has no extra term
that can cancel $\langle\mathbf{F}^O\rangle$. However, in this
case, the identity Eq. (\ref{hso1}) shows that
\begin{equation}
0=\frac{\partial^2\Delta_n(\mathbf{k})}{\partial k_i\partial
k_j}\frac{qm}{\hbar^2}E_j+\frac{qm}{\hbar^2}\Gamma^{n}_{ij}E_j,
\end{equation}
and thus, Eq. (\ref{hso4}) implies that
\begin{equation}\label{hso5}
\begin{split}
2Re\langle n\mathbf{k}|\frac{m}{\hbar^2i}[\frac{\partial
h_{so}}{\partial
k_i},h_{so}]|n\mathbf{k}\rangle^1&=-\frac{qm}{\hbar^2}\Gamma^{n}_{ij}E_j\\
&=\frac{\partial^2\Delta_n(\mathbf{k})}{\partial k_i\partial
k_j}\frac{qm}{\hbar^2}E_j.
\end{split}
\end{equation}
In general, $2Re\langle
n\mathbf{k}|\frac{m}{\hbar^2i}[\frac{\partial h_{so}}{\partial
k_i},h_{so}]|n\mathbf{k}\rangle^1$ does not equal
$\frac{\partial^2\Delta_n(\mathbf{k})}{\partial k_i\partial
k_j}\frac{qm}{\hbar^2}E_j$, as can be seen from Eq. (\ref{hso4}).
This important relation [Eq. (\ref{hso5})] shows that
$\langle\mathbf{F}^{O}\rangle$ still gives the mean SOIF
$\langle\mathbf{F}^{SOI}\rangle$ in a k-linear system. Therefore,
the total force in a k-linear system is still described by Eq.
(\ref{TotalF}). The validity of Eq. (\ref{hso5}) can be further
examined in the discussion of the Rashba-Dresselhaus system by the
calculation of $K_{ij}$ [Eq. (\ref{Kij})]. We will return to this
examination in the next section.

In the presence of an electric field, the force experienced by a
band electron (or hole) acts along the direction of the applied
electric field. However, when the band structure is anisotropic,
the force in the system may be perpendicular to the applied
electric field even in the absence of a magnetic field. The
existence of a transverse force can be confirmed by calculating
the off-diagonal component of $K_{ij}$ [Eq. (\ref{Kij})]. For
convenience, we assume that the external electric field is applied
in the +y direction (or [010] direction), and thus, the mean
transverse SOIF is $\langle F^{SOI}_x\rangle=qK_{xy}E_y$ and the
mean longitudinal SOIF is $\langle
F^{SOI}_y\rangle=q(n_c+K_{yy})E_y$. The integrand of Eq.
(\ref{Kij}) is of the form $\frac{\partial^2\Delta}{\partial
k_i\partial k_j}$, and we have the symmetric property:
$K_{ij}=K_{ji}$. If the energy dispersion is isotropic, such as in
the Rashba ($\Delta=\alpha k$)~\cite{Rashba84}, Dresselhaus
($\Delta=\beta k$)~\cite{Dress55}, k-cubic Rashba
($\Delta=\alpha_Rk^3$)~\cite{Winkler00}, and wurtzite
($\alpha_ok+\beta_ok^3$)~\cite{Zorkani96} systems, the quantity
$K_{xy}$ (and $K_{yx}$) always vanishes because
$\frac{\partial^2\Delta}{\partial k_x\partial
k_y}=k[\frac{\partial}{\partial
k}(\frac{1}{k}\frac{\partial\Delta}{\partial k})]\sin\phi\cos\phi$
and the integration of $\sin\phi\cos\phi$ from $0$ to $2\pi$ is
zero. For the spherical Luttinger Hamiltonian ($\Delta_{\pm
1/2}=2\gamma_2\frac{\hbar^2k^2}{2m_0}$ and $\Delta_{\pm
3/2}=-2\gamma_2\frac{\hbar^2k^2}{2m_0}$), all off-diagonal
components of $K_{ij}$ are zero because the dispersion in this
system is also isotropic. Therefore, the non-vanishing transverse
force can exist only in an anisotropic system.

However, it must be pointed out that not all the anisotropic
systems would have the non-vanishing mean transverse SOIF. For the
2D Dresselhaus type system along the [110]
direction~\cite{Zutic04}, the energy dispersion is $\sqrt{2}\rho
k_x$, and the second derivative of this energy dispersion leads to
zero. For the 2D k-cubic Dresselhaus system~\cite{Malshu05,
TWChen08}, the mean transverse SOIF also vanishes. The energy
dispersion of the 2D k-cubic Dresselhaus system is
$\Delta=\beta_Dkk_xk_y$, and we find that the band structure with
$n=+$ can be obtained from the other one with $n=-$ by using the
$\pi/2$ coordinate rotation, i.e., $k_x\rightarrow k_y$ and
$k_y\rightarrow -k_x$, and vice versa. On the Fermi surface, we
have
$\epsilon_F=\frac{\hbar^2k_n^2(\phi)}{2m}-n\beta_Dk_n^3(\phi)\sin\phi\cos\phi$,
and the right-hand side of this equality must be invariant under
$\phi\rightarrow -\phi$. Using the fact that the function
$\sin\phi\cos\phi$ is an odd function of $\phi$, we obtain
$\epsilon_F=\frac{\hbar^2k^2_+(-\phi)}{2m}+\beta_Dk^3_+(-\phi)\sin\phi\cos\phi=\frac{\hbar^2k^2_-(\phi)}{2m}+\beta_Dk^3_-(\phi)\sin\phi\cos\phi$,
and thus, we have $k_{+}(-\phi)=k_-(\phi)$. Furthermore, Eq.
(\ref{Kij}) can be rewritten as
$\int_{-\pi}^{\pi}d\phi(k_+^3-k_-^3)(2-\cos^2\phi\sin^2\phi)$, and
thus, it vanishes by using the result $k_{n}(-\phi)=k_{-n}(\phi)$.
The same conclusion can also be obtained by means of symmetry
consideration. Because the energy dispersion is invariant under
$\pi/2$ rotation around the z-axis, we obtain $K_{xx}=K_{yy}$ and
$K_{xy}=-K_{yx}$ by means of
$K(\frac{\pi}{2})=U(\frac{\pi}{2})KU^{\dag}(\frac{\pi}{2})$, where
$U(\frac{\pi}{2})$ is the 2$\times$2 rotation matrix around z-axis
by $\pi/2$. Due to the result that $K_{xy}=K_{yx}$, we have
$K_{xy}=0$. This implies that the mean transverse SOIF vanishes in
the 2D k-cubic Dresselhaus system and charge imbalance does not
occur in this case.

\begin{figure}
\begin{center}
\includegraphics[width=8cm,height=12.4cm]{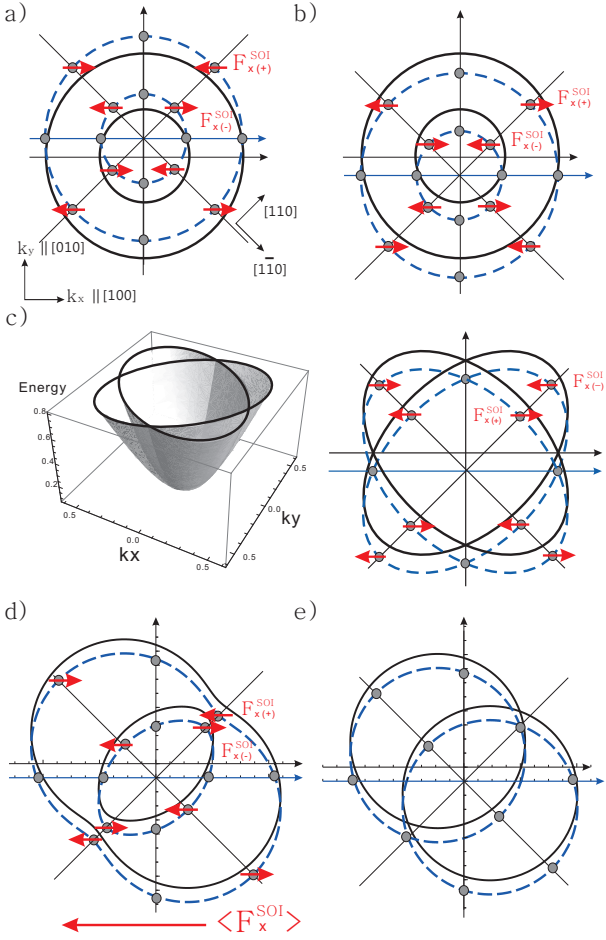}
\end{center}
\caption{(Color online) The transverse SOIF vector
($F^{SOI}_{x(n)}=-qn\frac{\partial\Delta}{\partial k_x\partial
k_y}|E_y|$) distribution. The arrow (red) represents the
transverse SOIF vector, and the dashed line (blue) represents the
deviation in the energy band due to the electric field applied in
the +y direction ([010] direction). (a) hole system with isotropic
dispersion; (b) electron system with isotropic dispersion; (c) 2D
k-cubic Dresselhaus system; (d) Rashba and Dresselhaus system with
$\alpha\neq\beta$; (e) Rashba and Dresselhaus system with
$\alpha=\beta$.}
\end{figure}

The SOIF vector distribution at a 2D $k$-plane is shown in Fig. 1,
where only the transverse SOIF is illustrated in the figure. The
electric field is applied in the $+y$ direction (i.e., $[010]$
direction), and thus, the energy dispersion would be displaced
toward $+y$ direction in hole system and toward $-y$ direction in
electron system. $F^{SOI}_{x(n)}=-qn\frac{\partial\Delta}{\partial
k_x\partial k_y}|E_y|$ represents the transverse SOIF of the
$n$-th band. The SOIF vector distribution in the hole isotropic
system is given in (a) and that in the electron system is given in
(b). The transverse SOIF vector of the 2D k-cubic Dresselhaus
system is shown in Fig. 1. (c). At the angles $0$,
$\frac{\pi}{2}$, $\pi$, and $\frac{3\pi}{2}$, the SOIF vanishes in
the isotropic and anisotropic systems. We note that in the
isotropic system, the electron (hole) concentration in the $k$-plane
is symmetric along the $y$-axis, and the mean transverse
SOIF is zero. This implies that the applied electric field would
not result in the charge imbalance at the edges of the sample. In
the Rashba-Dresselhaus system ((d) in Fig. 1.), the electron
concentrations on the left-hand and right-hand sides of the
$k$-plane are equal. However, the mean transverse SOIF does not vanish (see
the following section). In this system, the non-vanishing mean
transverse SOIF acts as the driving force and results in charge
imbalance at the edges of the sample. This implies that a
non-zero Hall voltage can be detected in the Rashba-dresselhaus
system when only an electric field is present. This is rather
different from the classical charge Hall effect, where an external
magnetic field must be applied to the system.

\section{Rashba-Dresselhaus system}
In this section, we study the mean SOIF in the Rashba-Dresselhaus
system. The Hamiltonian can be written as
$H_0=\frac{\hbar^2k^2}{2m}+\alpha(\sigma_xk_y-\sigma_yk_x)-\beta(\sigma_xk_x-\sigma_yk_y)$.
The energy dispersion is
$E_{nk}=\frac{\hbar^2k^2}{2m}-n\Delta(\mathbf{k})$, where
$\Delta(\mathbf{k})=\sqrt{(\alpha^2+\beta^2)k^2-4\alpha\beta
k_xk_y}=k\gamma(\phi)$, and
$\gamma(\phi)=\sqrt{\alpha^2+\beta^2-2\alpha\beta\sin(2\phi)}$. In
our notation, $\Delta_n(\mathbf{k})=-n\Delta$ and
$h_{so}=\alpha(\sigma_xk_y-\sigma_yk_x)-\beta(\sigma_xk_x-\sigma_yk_y)$.

\subsection{Non-vanishing longitudinal and transverse SOIF}
In the Rashba-Dresselhaus system, the longitudinal coefficient
$K_{yy}$ can be evaluated analytically:
\begin{equation}
K_{yy}=-\frac{m^2}{2\pi\hbar^4}|\alpha^2-\beta^2|.
\end{equation}
We note that $K_{yy}$ is always negative in the Rashba-Dresselhaus
system when $\alpha\neq\beta$. This means that the longitudinal
SOIF acts in a direction opposite to that of Lorentz force. The
longitudinal coefficient $K_{yy}$ is approximately $10^{-7}
(1/$\AA$^2)$ in semiconductor with large built-in electric field
($\alpha\sim 10^{-2}$eV\AA)~\cite{Cho05}, which is very small in
comparison with the value of the Lorentz force (free carrier
density is approximately $10^{-4} (1/$\AA$^2)$ in the
Rashba-Dresselhaus system, see Ref.~\cite{Gan06}). The mean
longitudinal SOIF is non-zero in the Rashba system ($\beta=0$) as
well as in the Dresselhaus system ($\alpha=0$). On the other hand,
the integrand of the off-diagonal component of $K_{ij}$
(transverse coefficient) is $\frac{\partial^2\Delta}{\partial
k_x\partial k_y}=-\frac{(\alpha^2-\beta^2)^2}{\Delta^3}k_xk_y$.
The transverse coefficient $K_{xy}$ can be evaluated analytically.
We find that $K_{xy}$ also depends on the Rashba and Dresselhaus
couplings, and in general, it does not vanish if
$\alpha\neq\beta$:
\begin{equation}\label{KijRandD}
\begin{split}
K_{xy}&=-\frac{m}{4\pi^2\hbar^2}\int^{k_+}_{k_-}dS_k\frac{\partial^2\Delta}{\partial
k_x\partial k_y}\\
&=\frac{m^2}{2\pi\hbar^4}\left\{\begin{array}{cc}
\displaystyle{\frac{\beta}{\alpha}|\alpha^2-\beta^2|}&~;~\alpha^2>\beta^2\\
\\
\displaystyle{\frac{\alpha}{\beta}|\alpha^2-\beta^2|}&~;~\beta^2>\alpha^2
\end{array}\right..
\end{split}
\end{equation}
Therefore, the system has a net SOIF in the -x direction, namely,
$\langle F_x^{SOI}\rangle=-|e|K_{xy}E_y$ and $K_{xy}>0$ (see Fig.1
(d)). When the Rashba coupling is equal to the Dresselhaus
coupling, the transverse SOIF vanishes, and further the energy
dispersion is not isotropic
($\Delta(\alpha=\beta)=\sqrt{2}\alpha|k_x-k_y|$; see Fig. 1. (e)).
%On the other hand, spin-Hall conductivity also vanishes when
%$\alpha=\beta$, and thus, charge imbalance does not occur in this
%case. This result is in agreement with the calculation of
%$K_{ij}$.

In the Rashba-Dresselhaus system, we have $\partial^2
h_{so}/\partial k_i\partial k_j=0$, and the relation in Eq.
(\ref{hso5}) follows. This can be examined by the direct
calculation of the conventional spin Hall conductivity. The force
$F_x^O=\frac{4m^2}{\hbar^4}(\alpha^2-\beta^2)(\mathbf{J}^{s_z}\times\hat{e}_z)_x=\frac{4m^2}{\hbar^4}(\alpha^2-\beta^2)J_y^{s_z}$
is shown to be proportional to the conventional spin longitudinal
conductivity $\sigma^s_{yy}$~\cite{Shen05}. It can be shown that
$\sigma^s_{yy}=\frac{q}{8\pi^2}(\alpha^2-\beta^2)\int_0^{2\pi}d\phi\frac{\sin\phi\cos\phi}{\gamma(\phi)^2}$.
We have $\sigma^s_{yy}=\frac{q}{8\pi}\frac{\beta}{\alpha}$ for
$\alpha^2>\beta^2$ and
$\sigma^s_{yy}=-\frac{q}{8\pi}\frac{\alpha}{\beta}$ for
$\beta^2>\alpha^2$. We obtain $\langle
F_x^O\rangle=\frac{qm^2}{2\pi\hbar^4}\frac{\beta}{\alpha}(\alpha^2-\beta^2)E_y$
for $\alpha^2>\beta^2$ and $\langle
F_x^O\rangle=\frac{qm^2}{2\pi\hbar^4}\frac{\alpha}{\beta}(\beta^2-\alpha^2)E_y$
for $\beta^2>\alpha^2$; this is in agreement with the result of
Eq. (\ref{KijRandD}). Therefore, physically the linear response of
$\mathbf{F}^{O}$ to the weak electric field is identical to the
semiclassical force acting on a band electron in this case. Due to
the non-vanishing transverse force in the Rashba-Dresselhaus
system, the charge imbalance would exist in the system even in the
absence of an external magnetic field because the mean transverse
SOIF acts as the driving force.

On the other hand, when an electric field is applied in the
$[\bar{1}10]$ direction denoted as $k_y'$ direction, the effective
Hamiltonian can be rewritten as
$H'_0=\frac{\hbar^2k'^2}{2m}+\frac{1}{\sqrt{2}}(\sigma_x-\sigma_y)(\alpha-\beta)k_x'+\frac{1}{\sqrt{2}}(\sigma_x+\sigma_y)(\alpha+\beta)k_y'$,
where the energy dispersion is
$\gamma'(\phi)=\sqrt{(\alpha-\beta)^2\cos^2\phi+(\alpha+\beta)^2\sin^2\phi}$.
In this case, it can be shown that $K_{x'y'}$ integral vanishes.
It is interesting to note that the conventional spin-Hall
conductivity in this case is not an universal
constant~\cite{Sinova04}. We find that the conventional spin-Hall
conductivity depends on the strength of spin-orbit coupling:
$\sigma^z_{x'y'}=\frac{q}{8\pi}(1+\frac{\beta}{\alpha})$,
$\alpha>\beta$ and
$\sigma^z_{x'y'}=-\frac{q}{8\pi}(1+\frac{\alpha}{\beta})$,
$\alpha<\beta$. Furthermore, we find that
$\sigma^{z}_{y'x'}=-\frac{q}{8\pi}(1-\frac{\beta}{\alpha})$,
$\alpha>\beta$ and
$\sigma^{z}_{y'x'}=\frac{q}{8\pi}(1-\frac{\alpha}{\beta})$,
$\beta>\alpha$, from which we have an unsymmetrical result
$\sigma^{z}_{x'y'}\neq-\sigma^{z}_{y'x'}$. This is because $[110]$
and $[1\bar{1}0]$ are nonequivalent axes in the sense that the
Rashba-Dresselhaus system has the $C_{2v}$ symmetry~\cite{Wink03}.
The resulting spin splittings along the $[110]$ and $[1\bar{1}0]$
directions are equal to $2(\alpha+\beta)k'$ and
$2(\alpha-\beta)k'$, respectively.

\subsection{The Hall voltage}

When an external in-plane electric field is applied to the
Rashba-Dresselhaus system, the charge accumulation generated by
the spin Hall current does not result in charge imbalance because
the system has equal populations of electrons with $+z$ and $-z$
polarized spins. However, as shown in the above subsection, we
find that the mean transverse SOIF which is independent of spin
components does not vanish in the Rashba-Dresselhaus system. As a
result, the existence of the non-zero transverse force leads to a
charge imbalance between two edges of the sample. Furthermore, we
can estimate the resulting Hall voltage from the classical point
of view. When the system reaches an equilibrium, the transverse
electric field generated by the charge imbalance can balance the
mean single-particle transverse SOIF which is defined as
\begin{equation}\label{SPSOIF}
\langle\langle F^{SOI}_x\rangle\rangle=\frac{\langle
F^{SOI}_x\rangle}{n_c},
\end{equation}
where
$n_c=\frac{1}{4\pi}\left[\frac{4m\epsilon_F}{\hbar^2}+(\frac{2m}{\hbar^2})^2(\alpha^2+\beta^2)\right]$
is the electron concentration. Therefore, an equilibrium state
requires $\langle\langle
F^{SOI}_x\rangle\rangle=qE_x=q\frac{V_H}{W}$, where $V_H$ is the
Hall voltage, and $W$ is the width of a device. On the other hand,
$\langle\langle
F^{SOI}_x\rangle\rangle=(K_{xy}/n_c)qE_y=(K_{xy}/n_c)q\frac{V_L}{L}$,
where $V_L$ is the longitudinal voltage and $L$ is the length of a
device. Therefore, the Hall voltage can be written as
\begin{equation}\label{VH}
V_H=\frac{K_{xy}}{n_c}\frac{W}{L}V_L.
\end{equation}
Unlike the classical charge-Hall effect, the non-zero Hall voltage
occurs in the absence of an external magnetic field. Figure 2(a)
shows $\frac{K_{xy}}{n_c}$ as a function of the Rashba coupling,
where the Dresselhaus coupling is fixed to $\beta=7.5\times
10^{-4} eV\AA$. As shown in Fig. 2(a), the maximum value of
$K_{xy}/n_c$ for the Fermi energy being smaller than $10^{-5}$ is
about $10^{-2}$. We assume that the length-to-width ratio is equal
to 8 and the longitudinal voltage is about $10^{-3}$
(Volt)~\cite{Kli80}. We find that the Hall voltage is nearly equal
to $1\mu V$.

Figure 1(d) shows the direction of the transverse SOIF vector in
different k points. If the mean transverse SOIF in $-x$ direction
is denoted as a minus sign and that in $+x$ direction is denoted
as plus sign, then the populations of electrons corresponding to
those different directions of the mean SOIF is illustrated in the
inset of Fig. 2(b). We define a deviation of electron
concentration as
\begin{equation}\label{deltan}
\delta n=n_{+}-n_{-},
\end{equation}
where $n_+$ and $n_-$ refer to the electron concentration with the
transverse SOIF vector in $+x$ and $-x$ directions, respectively.
In other words, $n_-$ is the sum over all shaded regions shown in
the inset of Fig. 2(b), and the sum over all unshaded regions
gives $n_+$. Each region ranges from zero to the corresponding
wave vector, and it follows that
\begin{equation}
\begin{split}
n_+&=\frac{1}{(2\pi)^2}\left(\int_0^{\pi/2}k_-^2(\phi)d\phi+\int^{\pi}_{\pi/2}k_+^2(\phi)d\phi\right)\\
n_-&=\frac{1}{(2\pi)^2}\left(\int_0^{\pi/2}k_+^2(\phi)d\phi+\int^{\pi}_{\pi/2}k_-^2(\phi)d\phi\right),
\end{split}
\end{equation}
where
$k_{\pm}(\phi)=\pm\frac{m}{\hbar^2}\gamma(\phi)+\left((\frac{m}{\hbar^2})^2\gamma(\phi)^2+\frac{2m\epsilon_F}{\hbar^2}\right)^{1/2}$.

\begin{figure}
\begin{center}
\includegraphics[width=8.6cm,height=12.6cm]{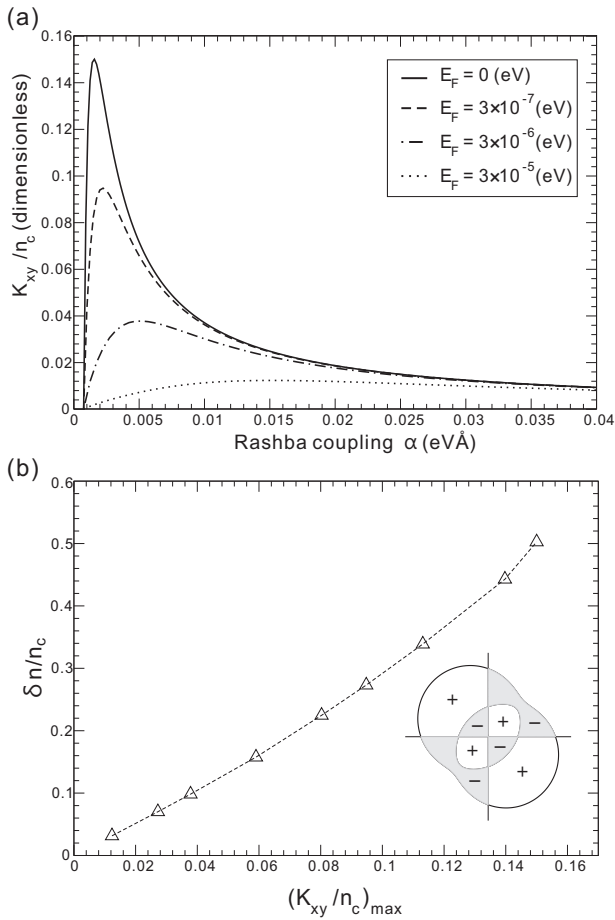}
\end{center}
\caption{(a) The dimensionless quantity $K_{xy}/n_c$ as a function
of the Rashba coupling; (b) The deviation of the electron
concentration $\delta n/n_c$ varies with respect to the maximum
value of $K_{xy}/n_c$.}
\end{figure}

Equation~(\ref{deltan}) is interesting because it is the origin of
charge imbalance that requires the population of electrons at two
edges of the sample to be different. In this case, we find that
the deviation $\delta n/n_c$ grows as the maximum value of
$K_{xy}/n_c$ denoted as $(K_{xy}/n_c)_{max}$ increases (see Fig.
2(b)). This result explains that the charge imbalance can be
characterized through the calculation of $K_{xy}$ and clarifies
the significance of our expression for the mean transverse SOIF.
Quantitatively, the calculation of $K_{ij}$ [Eq. (\ref{Kij})] is
important because the charge imbalance cannot reach to the value
$\delta n$, and thus, the resulting Hall voltage cannot be
determined from $\delta n$.

%The electrical measurement of intrinsic spin-Hall effect in the
%ballistic region has recently been carried out for HgTe
%nanostructures~\cite{Bru08}. We believe that the charge imbalance
%resulting from the non-zero transverse SOIF can be detected in the
%ballistic region by measuring the Hall voltage. Hopefully, the
%prediction would stimulate measurements of the Hall voltage in the
%Rashba-Dresselhaus system in the near future.

We close this subsection by some discussions on gauge
transformation and the charge Hall current. By using Eqs.
(\ref{LRT1}) and (\ref{LRT2}), the charge current $J_i=qv_i$ in
the ballistic region can be written as~\cite{MCChang05}
\begin{equation}\label{CHC}
\langle
J_i\rangle=\frac{q^2}{\hbar\mathcal{V}}\sum_{\mathbf{k}n}f_{n\mathbf{k}}\mathcal{F}^{(n)}_{ij}E_j,
\end{equation}
where
\begin{equation}\label{Fij}
\mathcal{F}^{(n)}_{ij}=\frac{\partial A^{(n)}_{j}}{\partial
k_i}-\frac{\partial A^{(n)}_{i}}{\partial k_j}
\end{equation}
is the field strength of $n$-th band in $k$-space, and it
satisfies the antisymmetric property
$\mathcal{F}^{(n)}_{ij}=-\mathcal{F}^{(n)}_{ji}$. The Berry vector
potential in Eq. (\ref{Fij}) is defined as $A^{(n)}_{i}=\langle
n\mathbf{k}|(-i)\frac{\partial}{\partial k_i}|n\mathbf{k}\rangle$.
We note that the longitudinal charge current always vanishes, as
shown in Eq. (\ref{Fij}), where the anti-symmetric property of
$\mathcal{F}^{(n)}_{ij}$ guarantees that $\mathcal{F}^{(n)}_{ii}$
is zero, and this result is independent of any gauge choice.
Physically, this means that there is no steady longitudinal
current without considering the collision with impurity. In that
sense, the existence of a steady charge-Hall current generated by
the spin Hall effect is possible because there is no transverse
electric field. However, the charge Hall current can not be
uniquely determined from Eq. (\ref{CHC}) because the singular
gauge transformation changes the magnitude of the charge Hall
conductivity in 2D spin-orbit coupled systems~\cite{MCChang05}. On
the other hand, the system has equal populations of electrons with
$+z$ and $-z$ polarized spins. Therefore, in
Ref.~\cite{MCChang05}, a specific gauge is chosen such that the
steady charge-Hall current is zero.

Nevertheless, the mean transverse SOIF does not vanishes in the
ballistic region, and thus, the non-steady charge Hall current
exists in the Rashba-Dresselhaus system. As a result, the charge
imbalance between two edges of the sample can occur in the system.
The system can still have an equilibrium state in the sense that
the transverse electric field generated by charge imbalance can
balance the mean single-particle transverse SOIF.

\subsection{Disorder effect}
In this section, we will consider the weak disorder effect. We
will show that the transverse SOIF vanishes if the calculation is
carried out to the order of $\frac{\Delta_{F}}{\epsilon_F}$ in the
sense that the Fermi energy ($\epsilon_F\equiv\hbar^2k_F^2/2m$) is
much larger than the spin-splitting ($\Delta_F=2\gamma(\phi)k_F$).
In the Rashba-Dresselhaus system, the transverse force is
determined by $\mathbf{F}^O$ because $\mathbf{F}^E=0$. It can be
represented by the term proportional to spin current
$J^z_y=\frac{1}{2}\{v_y,s_z\}$, i.e.,
$F_x^{O}=\frac{4m^2}{\hbar^4}(\alpha^2-\beta^2)J^z_{y}$. Therefore
the calculation of force is simplified to the calculation of
conventional spin-Hall conductivity. Because $F_x^O$ is
proportional to $J_y^z$, we have to calculate the
spin-longitudinal conductivity $\sigma^{s_z}_{yy}$.

The impurity is assumed to have a $\delta$-potential
$V(\mathbf{x})=V_0\mathbf{1}\sum_{i}\delta(\mathbf{x}-\mathbf{R}_i)$,
where $\mathbf{R}_i$ is the location of the $i$-th impurity and
$V_0$ is the potential strength. The relaxation time $\tau$ can be
obtained from the calculation of the self-energy by using the
definition $\frac{\hbar}{\tau}=2|Im\Sigma|$. For the impurity
potential
$V(\mathbf{x})=V_0\mathbf{1}\sum_{i}\delta(\mathbf{x}-\mathbf{R}_i)$,
the self-energy can be calculated by the the following
equation~\cite{Dug05}: $\Sigma(\epsilon)=\frac{\langle
n_iV_0^2\rangle}{2}\sum_s\int d\mathbf{k}G_{k,s}(\epsilon)$, where
$G_{k,s}(\epsilon)=1/[\epsilon-\epsilon_{ks}-\Sigma(\epsilon)]$,
and $n_i$ is the impurity concentration. In the Born
approximation~\cite{Dug05},
\begin{equation}
\frac{\hbar}{\tau}=\pi\langle n_iV_0^2\rangle(N_++N_-),
\end{equation}
where $N_{s}$ is the density of states of band $s$. In the
Rashba-Dresselhaus system, the total density of states is
$(N_++N_-)=\frac{1}{2\pi}(k_+\frac{dk_+}{d\epsilon_F}+k_-\frac{dk_-}{d\epsilon_F})=\frac{m}{\pi\hbar^2}$,
and thus, $\langle n_iV_0^2\rangle=\frac{\hbar^3}{m\tau}$. The
retarded and advanced Green's functions are diagonal in the
helicity space, that is,
\begin{equation}
G^{R,A}(\epsilon)=\left(
\begin{array}{cc}
  G^{R,A}_{+}(\epsilon)&0\\
  0&G^{R,A}_{-}(\epsilon)\\
\end{array}\right),
\end{equation}
where
\begin{subequations}
\begin{align}
G^{R}_s(\epsilon)&=\frac{1}{\epsilon-\epsilon_{ks}+i\Gamma}~;\\
G^{A}_s(\epsilon)&=\frac{1}{\epsilon-\epsilon_{ks}-i\Gamma}.
\end{align}
\end{subequations}
$\Gamma$ is defined as $\Gamma=\frac{\hbar}{2\tau}$. The spin
longitudinal conductivity $\sigma^{s_z}_{yy}$ can be evaluated by
calculating the following Green's function~\cite{Arii07}:
\begin{equation}\label{VerCorG}
\Theta^{RA}(\epsilon,\epsilon')=\frac{q\hbar}{2\pi}\int
d\mathbf{k}Tr[J_y^{z}G^{R}(\epsilon)\mathcal{V}_yG^{A}(\epsilon')],
\end{equation}
where $\epsilon'=\epsilon-\hbar\omega$ and $\mathcal{V}_y$
includes the vertex correction [see Eq. (\ref{VerCor})]. The
notation $Tr$ represents the trace in the helicity space $s$.
Here, the electric field is applied in the y-direction, and we
calculate the spin current in the y direction, which has a spin
component in the z direction. The spin longitudinal conductivity
$\sigma^{s_z}_{yy}$ can be evaluated by using the
formula~\cite{Arii07}
\begin{equation}
\sigma^{s_z}_{yy}=-\int
d\epsilon\frac{f(\epsilon-\hbar\omega)-f(\epsilon)}{\hbar\omega}\Theta^{RA}(\epsilon,\epsilon-\hbar\omega),
\end{equation}
where $f(\epsilon)$ is the Fermi-Dirac distribution. In the ladder
approximation, the vertex correction of electric velocity
$\mathcal{V}_x$ satisfies the self-consistent vertex
equation~\cite{Arii07}:
\begin{equation}\label{VerCor}
\mathcal{V}_y=v_y+\langle
n_iV_0^2\rangle\int\frac{d\mathbf{k}}{(2\pi)^2}G^{R}\mathcal{V}_yG^{A}.
\end{equation}
The electron velocity $\mathcal{V}_y$ can be divided into two
parts, $\mathcal{V}_y=v_y+\tilde{v}_y$, where $v_y$ is the
electron velocity in the absence of impurities and $\tilde{v}_y$
is the vertex correction in the presence of impurities. The vertex
correction of velocity, $\tilde{v}_y$, can be written as
$\tilde{v}_y=\sum_{\mu=0}^{3}c_{\mu}\sigma_{\mu}$ (See also Eq.
(\ref{GSofVelo})), and the general solution of Eq. (\ref{VerCor})
is given by~\cite{zhou07}
\begin{equation}\label{GSofVelo}
\mathcal{V}_y=v_y+\sum_{\mu}\sigma_{\mu}c_{\mu}~;~\mu=0,1,2,3,.
\end{equation}
The matrix $\sigma_0$ is the two by two identity matrix, and
$\sigma_i$ ($i=1,2,3$ corresponds to $x,y,z$) are the Pauli
matrices. We substitute Eq. (\ref{GSofVelo}) into Eq.
(\ref{VerCor}) and obtain the matrix equation
\begin{equation}\label{Qmn}
\sum_{\nu}Q_{\mu\nu}c_{\nu}=D_{\mu}~;~\mu, \nu=0,1,2,3,
\end{equation}
where
\begin{equation}
Q_{\mu\nu}=2\delta_{\mu\nu}-\langle
n_iV_0^2\rangle\int\frac{d\mathbf{k}}{(2\pi)^2}Tr[G^R\sigma_{\nu}G^{A}\sigma_{\mu}],
\end{equation}
where it can be shown that $Q_{30}=Q_{03}=0$ and
\begin{equation}
D_{\mu}=\langle
n_iV_0^2\rangle\int\frac{d\mathbf{k}}{(2\pi)^2}Tr[G^Rv_yG^{A}\sigma_{\mu}].
\end{equation}
It can be shown that $Q_{13}$, $Q_{23}$ (and thus $Q_{31}$ and
$Q_{32}$), and $D_3$ represent purely interband transitions. If we
assume that the scattering effect does not cause any interband
transition (i.e. this assumption is equivalent to
$\Delta_F>>\hbar\omega$), we can neglect these terms in the
calculation. Then in this case, the coefficient $c_z$ is zero. On
the other hand, if we restrict each $Q_{ij}$ and $D_j$ to the
order of $\Delta_{F}/\epsilon_F$, we can neglect $Q_{01}$,
$Q_{02}$ (as well as $Q_{10}$ and $Q_{20}$), and $D_0$ from which
we can conclude that $c_0=0$. Therefore, Eq. (\ref{Qmn})
simplifies to the following two by two matrix equation:
\begin{equation}\label{Qij}
\left(\begin{array}{cc}
Q_{11}&Q_{12}\\
Q_{21}&Q_{22}
\end{array}\right)
\left(\begin{array}{c}
c_x\\
c_y
\end{array}\right)
=\left(\begin{array}{c}
 D_1\\
D_2
\end{array}\right).
\end{equation}
In the DC limit $\omega\rightarrow 0$ (taken before the limit
$\tau\rightarrow\infty$), we obtain
$D_1=\frac{\alpha}{\hbar}(f_1-2)-\frac{\beta}{\hbar}f_2$,
$D_2=-\frac{\alpha}{\hbar}f_2+\frac{\beta}{\hbar}(f_3-2)$,
$Q_{11}=2-f_1$, $Q_{22}=2-f_3$, and $Q_{12}=Q_{21}=f_2$, where
$f_i$ $(i=1,2,3)$ are the following integrals:
\begin{equation}
\begin{split}
f_1&=\frac{1}{\pi}\int_0^{2\pi}
d\phi\frac{(\alpha\sin\phi-\beta\cos\phi)^2}{\gamma(\phi)^2},\\
f_2&=\frac{1}{\pi}\int_0^{2\pi}
d\phi\frac{(\alpha\sin\phi-\beta\cos\phi)(\alpha\cos\phi-\beta\sin\phi)}{\gamma(\phi)^2},\\
f_3&=\frac{1}{\pi}\int_0^{2\pi}
d\phi\frac{(\alpha\cos\phi-\beta\sin\phi)^2}{\gamma(\phi)^2};
\end{split}
\end{equation}
here,
$\gamma(\phi)=\sqrt{\alpha^2+\beta^2-2\alpha\beta\sin(2\phi)}$.
After integration, we have $f_1=f_3=1$ and
$f_2=-\frac{\beta}{\alpha}$ for $\alpha>\beta$, and
$f_2=-\frac{\alpha}{\beta}$ for $\alpha<\beta$. Substituting
$f_1$, $f_2$, and $f_3$ into Eq. (\ref{Qij}), we obtain $c_x$ and
$c_y$:
\begin{equation}\label{CxCy}
\left(\begin{array}{c}
c_x(\omega=0)\\
c_y(\omega=0)
\end{array}\right)=
\left(\begin{array}{c}
-\alpha/\hbar\\
-\beta/\hbar
\end{array}\right)~;~\alpha\neq\beta.
\end{equation}
However, we have
$\mathcal{V}_y=v_y+c_x\sigma_x+c_y\sigma_y=v_y^0+(c_x+\frac{\alpha}{\hbar})\sigma_x+(c_y+\frac{\beta}{\hbar})\sigma_y$,
where $v^0_y=\hbar k_y/m$. Therefore, the vertex correction
cancels the anomalous velocity
$\frac{\alpha}{\hbar}\sigma_x+\frac{\beta}{\hbar}\sigma_y$.
Finally, the spin longitudinal conductivity $\sigma^{s_z}_{yy}$ is
proportional to the term
$v_y^0v_y^0Tr[\sigma_zG^RG^A]=v_y^0v_y^0Tr[\sigma_zG^R\sigma_0G^A]$.
It can be shown that the trace directly equals zero, that is,
$Tr[\sigma_zG^R\sigma_0G^A]=0$, and thus, the spin longitudinal
conductivity is zero. This implies that the transverse force in
the Rashba-Dresselhaus system vanishes when the disorder effect is
considered. In the same approximation, the spin-Hall conductivity
also vanishes. This can be seen as follows. The spin-Hall current
is proportional to $k_x\sigma_z$, and the spin Hall conductivity
$\sigma^{x}_{xy}$ is proportional to the term
$v^0_yv^0_xTr[\sigma_zG^R\sigma_0G^A]$ in which the trace term is
the same as the spin longitudinal conductivity, and thus, we
obtain the vanishing spin-Hall conductivity. The result obtained
from Eq. (\ref{CxCy}) agrees with the expected result for the pure
Rashba system~\cite{Mish04,Cha05,Bry06,Arii07,zhou07} and pure
Dresselhaus system~\cite{Malshu05}. The influence of impurities on
spin-Hall transport in the Rashba and Dresselhaus system also has
been studied in Refs.~\cite{Loss03} and~\cite{syn04} by means of
Boltzmann equation and Kubo formula, respectively. However, the
vertex correction [Eq. (\ref{VerCor})] which is non-zero in the
diffusive regime~\cite{Cha05, Bry06} is not taken into account in
both calculations.

In short, Eq. (\ref{TotalF}) is valid in any spin-orbit coupled
system with the Hamiltonian given by Eq. (\ref{Ham}). The charge
imbalance would occur in the Rashba-Dresselhaus system in the
clean limit when the the electric field is applied in the system.
However, it is shown that the disorder can cancel the transverse
force, and thus, the charge imbalance would not occur in this
case.

%We must emphasize that Eq. (\ref{TotalF}) is valid in any
%spin-orbit coupled system with Hamiltonian Eq. (\ref{Ham}). In
%this paper, we only show that in Rashba-Dresselhaus system the
%transverse SOIF does not vanishes but the disorder cancels this
%transverse SOIF. There might be other system that has non-zero
%transverse force and the disorder effect does not cancel
%transverse force. In that case, the charge imbalance can also take
%place in the absence of magnetic field.

\section{Conclusions}
%%%%%%%%%%%%%%%%%%%%%%%%%%%% Conclusions %%%%%%%%%%%%%%%%%%%%%%%%%
In conclusion, we show that the mean SOIF (spin-orbit interaction
induced force, Eq. (\ref{hsoEO})) in any spin-orbit coupled
systems is determined by the second derivative of energy dispersion
[Eq. (\ref{Kij})] only.
%the mean SOIF determined in this manner is valid in any spin-orbit coupled systems.
The mean transverse SOIF acts as the driving force and
thus results in transverse charge imbalance in the spin-orbit coupled system.
It has been shown that the mean transverse SOIF vanishes in
rotationally invariant systems such as Luttinger, k-linear Rashba,
k-cubic Rashba, k-linear Dresselhaus, and wurtzite systems.
Furthermore, we find that the mean SOIF also vanishes in the 2D
k-cubic Dresselhaus system, where the energy dispersion is
anisotropic. Nonetheless, we find that the mean transverse SOIF
does not vanish in the Rashba-Dresselhaus system. This result can
be verified by measuring the Hall voltage in the absence of an
external magnetic field. The estimated magnitude of Hall voltage
is nearly equal to $1\mu$V when a longitudinal voltage of 1meV
is applied. In the presence of weak disorder, the
SOIF vanishes in the Rashba-Dresselhaus system because the
anomalous velocity and vertex correction accidently cancel each
other.

The electrical measurement of intrinsic spin-Hall effect in the
ballistic region has recently been carried out for HgTe
nanostructures~\cite{Bru08}. Therefore, we believe that the transverse
charge imbalance resulting from the non-zero transverse SOIF
can be detected in the ballistic region by measuring the Hall voltage.
Hopefully, our interesting prediction would stimulate measurements
of the Hall voltage in such spin-orbit coupled systems with an anisotropic
dispersion as the Rashba-Dresselhaus system in the near future.

\section*{ACKNOWLEDGEMENTS}
%%%%%%%%%%%%%%%%%% ACKNOWLEDGEMENTS %%%%%%%%%%%%%%%%%%%%%%%%%%%%%%%%%%%%%%
The authors thank M. C. Chang and C. D. Hu for valuable
discussions. The authors gratefully acknowledge financial support
from the National Science Council and NCTS of Taiwan.

\end{document}